# Next-generation Web Applications with WebAssembly and TruffleWasm

M. Šipek, D. Muharemagić, B. Mihaljević, and A. Radovan
Rochester Institute of Technology Croatia, Zagreb, Croatia
matija.sipek@mail.rit.edu, dino.muharemagic@mail.rit.edu,
branko.mihaljevic@croatia.rit.edu, aleksander.radovan@croatia.rit.edu

*Abstract* - In modern software development, the JavaScript ecosystem of various frameworks and libraries used to develop contemporary web applications presents many advantages. JavaScript is a widely known interpreted programming language, simple to learn and start development, and with numerous third-party libraries and extensions. However, with the rise of highly user-interactive websites and browser-based games, in some cases, JavaScript's executable engine could lack in performance. Therefore, developers could combine several other programming languages to create a polyglot user-interactive interoperable system to develop efficient modern web applications. The interoperability modules offer significant advantages but also present challenges in the execution due to high complexity and longer compilation times.

This paper explores WebAssembly, a binary format compilation target with a low-level assembly-like language used for targeting from other programming languages. The binary format allows near-native performance level due to its compactness, as it prioritizes usage of low-level languages. Moreover, as a continuation of our previous research of the GraalVM ecosystem, we analyzed a guest language implementation of a WebAssembly based system, TruffleWasm, hosted on GraalVM and Truffle Java framework. This paper presents the architecture and review of the TruffleWasm within the GraalVM-based ecosystem as well as from performance test results within our academic environment.

## I. Introduction

Web browsers are currently the most omnipresent platform for user-interactive applications since they are used on a wide range of operating systems (OS) and devices, including desktops, laptops, tablets, and smartphones. Together with HTML and CSS, JavaScript has been the core built-in language implementation, which is supported by all major web browsers, including Google Chrome, Safari, Firefox, Microsoft Edge, and Brave. JavaScript is the primary language natively supported on web; thus, it has been the *de facto* standard programming language in web-based ecosystems with unparalleled usage compared to other technologies. Yet, as users expect highly interactive *progressive web-applications* (PWA), advanced web-based games or systems, JavaScript by itself, often lacks sufficient resources. To solve these challenges, developers are often forced to use multiple programming languages, thus creating a polyglot environment in order to make such a system. While interoperability offers many additional application capabilities, the result is an unnecessarily complex system, which, in the end, is not as usable as traditional, JavaScript-based applications.

However, JavaScript provides many functionally redundant libraries and optimizations; most of the optimizations affect less than 25% of code, more than 47 % of optimizations do not affect the cyclomatic complexity of the code, hardly 40% of optimizations provide stable performance enhancements, and even some 15% report a degrade of performance [1]. Alongside this, systems written in JavaScript often run slower than their native corresponding systems, since JavaScript is quite challenging to compile efficiently [2].

Nevertheless, JavaScript has been used as a compilation target for other programming languages because of its ubiquity. One example is a statically typed sublanguage called asm.js[1], a stylized subset of JavaScript used as a low-level target language for compilers. Afterward, this code can be optimized by LLVM [2] and pushed through a compiler toolchain Emscripten[3], the first C++ to JavaScript compiler. Again, the resulting system has inconsistent efficiency, a property frequently labeled on a scripting language-based system as denoting problems with dynamic type checks, garbage collections, and unpredictable intercommunication with the JIT compiler.

Additionally, there have been efforts to make add-on browser extensions, standards, and packages that tried to emulate low-level code's operability and performance; popular examples would be Microsoft's ActiveX, Native Client (NaCl), and Adobe Shockwave Player. These products work in their own respective fields as a Minimum Viable Product (MVP), yet, due to lack of integrating capabilities, portability, and security problems, they mostly became deprecated.

Consequently, a group of researchers from major browser vendors tried to fix these problems by creating WebAssembly[4], a portable low-level binary format that aims to run applications in a browser with performance similar to native applications. At the core, WebAssembly acts like a stack-based virtual machine. More specifically, it's a *virtual instruction set architecture* (virtual ISA), meaning its computational model can be used as a compilation target for other programming languages. The result is an executable, versionless, feature-proof, backward-compatible, and easy to integrate into the existing open web platform [3].

---

[1] asm.js, http://asmjs.org/spec/latest/
[2] LLVM, https://llvm.org/
[3] Emscripten, https://emscripten.org/index.html
[4] WebAssembly, https://webassembly.org/

Due to our previous research [4][5], we additionally investigated an overlapping point of interest conjointly with WebAssembly, which is TruffleWasm, a WebAssembly guest language implementation that uses underlying technologies such as GraalVM[5] and Truffle Framework. Truffle is a language implementation Java framework for creating self-optimizing *abstract syntax tree* (AST) based interpreters, which represents a program that can be run on a standard *Java Virtual Machine* (JVM). In essence, GraalVM is a high-performance polyglot JVM, but its modular design allows natural polyglotism within the system, meaning Truffle supported languages can reuse WebAssembly components and libraries. It is important to stress that TruffleWasm isn't an attempt to create a new runtime, as, for example, Wasmtime, but rather as a WebAssembly interpreter implementation [6].

Standalone WebAssembly modules need a system interface in order to access external resources during runtime, for example, files, clocks, random numbers, and Berkeley sockets. One approach is the *WebAssembly Systems Interface* (WASI), an Application Programming Interface (API) allowing WebAssembly to interact with external systems [7]. WebAssembly cannot use LLVM bitcode as LLVM compiler infrastructure for *Instruction Set Architecture* (ISA) and binary encoding as it differs from WebAssembly required standards. This results in an opportunity to plug in some other deployment-specific compiler to maximize performance.

Furthermore, in order to support the performance of polyglot applications, such an environment needs a modularly designed back-end; namely, an optimizing JIT compiler Graal endorsing aggressive speculative optimizations and AST specific transformation for Truffle guest languages. Although, as mentioned in the introduction, polyglotism can infer to a slow and complex system, GraalVM gives a highly optimized runtime that supports the direct execution of standalone WebAssembly environments. Currently, according to CanIUse [6], WebAssembly is supported by over 92% of browser globally, with production deployments in the current version of all major browsers:

- Edge versions 16-87 and 88
- Firefox versions 52, 53-84, 85, and 86-87
- Chrome versions 57-87, 88, and 89-91
- Safari versions 11-13.7, 14.4, and TP
- Opera versions 44-72 and 73

Supported mobile browsers include iOS Safari, Android browser, Chrome, and Firefox for Android.

Lastly, we want to mention a similar project called GraalWasm[7], an open-source project able to interpret and compile WebAssembly and implement it on the GraalVM. However, Graal doesn't support WASI and has not been tested thoroughly enough with only micro-benchmark presented; it would be no exaggeration to say that it is still unfeasible and in the early development phase.

## II. WebAssembly Ecosystem Analysis

### A. Background and Related Work

It is well known that World Wide Web started as a relatively simple exchange network, and soon enough, the lack of capacity for the dynamic behavior of web sites erupted a need for a new language. The initial idea was to use Java, a rich development ecosystem, but it was quite memory-heavy and mainly used by professional programmers. Thus, a lightweight interpreted language JavaScript was created, which lead the way for easier further development. Due to historical circumstances and the plainness of the code, the developer community continued with the expansion of the project, making it scalable, fast, and standardized.

In addition, the introduction of AJAX technology allowed to asynchronously affect the web site without reloading, pushed JavaScript usage to new heights; and, today, it is still the primary native language supported in the web ecosystem [8]. Unfortunately, due to JavaScript design, it is becoming unfeasible to run programs with large numerical computations that are highly CPU demanding. Nowadays, JavaScript returns a performance up to two times slower compared to native C/C++ applications, as it is becoming increasingly difficult to find areas of optimization [9].

WebAssembly's predecessors opened the door for the realization of low-level code on the web, such as Emscripten, an LLVM-to-JavaScript compiler, and asm.js, and testing results showed that it is possible to achieve near-native speed in the browser [10].

Additionally, as computing power saw a significant increase in mobile and Internet of Things (IoT) devices, it is important and necessary to use that power in order to increase user privacy [12]. The low-level disposition of WebAssembly provides many advantages for these kinds of optimizations and features:

- Effective encryption algorithms for numerically-intense security features
- Machine learning models and Big data computations
- Streamable and parallelizable
- Runtime environment (RE) embeddable
- Standalone runtime capable

---

[5] GraalVM, https://www.graalvm.org/
[6] CanIUse, https://caniuse.com/?search=WebAssembly
[7] GraalWasm, https://github.com/oracle/graal/tree/master/wasm

*B. Binary Format and Intermediary Compiler Target*

When comparing JavaScript and WebAssembly compilation processes, JavaScript transforms source code to AST before turning it to binary format, whereas WebAssembly skips that step and decodes source code directly to binary. Moreover, WebAssembly is a statically typed language removing the need for optimizations like type speculation during compilation time. However, WebAssembly lacks a memory management system like C/C++ language, which increases ambiguity; nevertheless, if used correctly, it increases the security and performance of the given system.

If we follow the official definition, WebAssembly is a low-level instruction format transmitted via the network as a binary encoding. Despite the name, WebAssembly is neither strictly web-based nor an assembly language, as it is not meant to run on a specific machine or environment but rather runs on any host system as a virtual ISA [12].

An important concept is a WebAssembly module, a unit of deployment, loading, and compilation, which holds definitions for functions, types, tables, global values, and other auxiliary data [3]. This binary encoding represents a single module separated into segments with different entities and additional sections. Notably, the code for functions calls is set in these separated segments, and it allows *streaming compilation*; the browser is already starting compiling function bodies before all of the packets have been received. The compilation process can be further speed up by *parallelize* compilation, which pushes the compilation speed up to a substantial result of 10-15 times faster [3][12].

Design of WebAssembly attentions to portability, also making it an intermediary compiler target as it is:

- Hardware-independent,
- Language-independent, and
- Platform-independent

Prior attempts were not targeting multiple environments on multiple levels but rather a single architecture, thus inducing a low portability system.

Listed key features facilitate further development, extensibility, and flexibility of the WebAssembly ecosystem. Similarly, the web's ecosystem goals overlap with it, as it extends over many different OS since WebAssembly aims to support various languages, programming modules, and object modules. Moreover, it can be compiled on any modern hardware machine architectures, desktop, IoT, or mobile device. Interoperability with WebAssembly environment is universal as it can be integrated into browsers, run as a standalone, or embedded in other host environments.

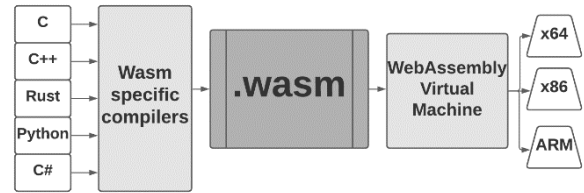

Figure 1. WebAssembly Execution Process

*C. Internal System Design*

Internal System design is rather complex, as it embodies all of the key structures and goals mentioned; thus, we will only cover the basic features and emphasize vital aspects. Even though WebAssembly is a binary code format, it is also a language with syntax and structure, and many more possibilities.

As presented in Fig. 1, developers do not write code in WebAssembly, since it is an instruction set, more precisely, an intermediate representation (IR) for the WebAssembly Virtual Machine (WAVM), comparable to bytecode in Java or common intermediate language (CIL) in .NET. The development process starts from known programming languages, and currently, there are several languages stable for production usage, including .NET, AssemblyScript, C, C#, C++, and Go.

Furthermore, in order to transform the source code into a WASM file, we need to use some kind of a WebAssembly specific compiler. It is important to notice that there is no such thing as a "silver bullet" in this aspect, but due to a modular system design, there are several options developers can utilize. Seasoned LLVM, which can be generated from C/C++ and pushed through Emscripten, and rustc, the Rust compiler, can generate WebAssembly directly. In addition, there are tools such as Cheerp, Blazor, and WasmFiddle, which all allow the compilation of virtually any C/C++ code to WASM, asm.js, JavaScript, or a combination of given.

Finally, the code is forwarded to a particular WebAssembly Virtual Machine, which turns IR into a platform-specific Machine Code and executes it. The VM's portability stretches the functionality further than only as a web-based, but rather in conjuncture with a web-based ecosystem. This flexibility allows the same code to be run on MacOS, Windows, Linux, or Android platforms, using different processing architectures of x86, x64, or ARM.

*D. WebAssembly Semantics*

WebAssembly has underlying core features that often seem in similar stack machines such as Modules, Instructions, Traps, Functions, and Machine Types. The main optimizations begin with Linear Memory which is a large byte array. There are several functionalities you can that can be performed here:

- *Creation and Growing* instructions allowing sharing of import/export data between modules,

- *Access* instructions provide access to memory via packed integer loads, 8-,16-,32- or 64-bit store or sign extensions,

- *Endianness* defines WebAssembly memory as little-endian byte order, meaning the least significant bytes in the list is stored first, and new digits are added on the right to the highest address

This makes memory access deterministic and thus portable across all platforms and engines. Subsequently, the Linear Memory is detached from attackable surfaces: code space, data structures located in engine modules, the execution stack; thus, disallowing corrupt program execution and undefined behavior [3].

WebAssembly code is validated, compiled, or transformed to intermediate form in a single pass due to its *structured control flow,* analogous to a programming language. This removes the possibility of irreducible loops or having branches in unfitting memory places such as the middle of a multi-byte instruction [3]. By design, features such as *Control Constructs and Blocks*, *Branches and Labels*, and *Expressiveness* provide a structured text format easier to read by developers.

Furthermore, the execution of code on the web pulled from unverified sources needs to be validated in order to be executed safely. The Validation section is in charge of security enforces key concepts [13]:

- *Typing Rules* for function types, polymorphic instructions, and modules,
- *Soundness* which enforces all execution states and rules to ensure the validity of programs.

### E. Interoperability and System Implementation

The WebAssembly abstraction over hardware and the separation of concerns allows developers to map their required data types to the memory, meaning that any language or object model can be implemented without fear of incompatibility.

Implementation of WebAssembly to JavaScript engines in major browsers achieves very high-performance results without forfeiting portability as V8, Spider-Monkey and JavaScriptCore reuse their JIT compiler for Ahead-of-Time compilation (AoT). This removes system response randomness during warmup time and ensures faster startup time and lower memory spending.

### III. TRUFFLEWASM INTERPRETER OVERVIEW

#### A. Polyglot Support to WebAssembly Ecosystem

Contemporary JVM presents a highly optimized runtime that has been tested thoroughly and deployed to various platforms. The GraalVM presents a JVM distribution written in Java, meaning easier development and maintenance than the standard JVM. GraalVM architecture offers many possibilities as it is modular and stacked; it consists of the JVM compiler interface (JVMCI), which allows the implementation of a custom optimizing dynamic compiler; in our case, the Graal JIT compiler. Further, the Truffle framework enables implementing multiple programming languages for the Graal compiler via self-optimizing interpreters. It uses an AST to incorporate a particular version of an interpreter in order to compile the source of a *guest* programming language.

#### B. TruffleWasm

The TruffleWasm ecosystem starts with front-end compilers such as Emscripten, Rust Compiler, or others, before being transformed to WASM. This a similar sequence as compared to the standard WebAssembly compilation process. However, in order to connect the WASM binaries, TruffleWasm is using *Binaryen* [8], a compiler infrastructure and toolchain library used for parsing and validation. The WebAssembly IR is transformed into tree-based Binaryen IR, which can be further adapted into Truffle AST for interpretation by the GraalVM [14].

To maximize performance, TruffleWassm needs to implement WASI functions, as WebAssembly requires a standalone runtime. These functions are called using C++ wrapper functions via POSIX API, and check all necessary configurations and requirements [15].

#### C. Interoperability with GraalVM languages

TruffleWasm reuses GraalVM components via the Truffle framework to generate highly optimized code. To access libraries and other framework accessories provided by WebAssembly, GraalVM languages have to go through C/C++ and Rust modules. Moreover, these languages can communicate directly with the WASI API, which reduces the need for TruffleWasm to maintain other language-specific details related to objects and functions in the current context.

### IV. BENCHMARK AND RESULTS

#### A. Benchmark Test Cases

Selecting the correct set of benchmark tests is crucial to showcase the benefit of WebAssembly in real-world applications. Thus, for this research, we selected a variable set of computationally-intensive workloads that demonstrate the best how WebAssembly alleviates bottlenecks that often occur in pure JavaScript code implementations. The following benchmarks have been included in our analysis:

- *Fibonacci* – a standard algorithm implementation that calculates the Fibonacci sequence for term *n*,
- *Collision detection* – an algorithm that performs collision detection of simple 2D shapes,
- *MultiplyIntVec* – multiplies two integer vectors,
- *QuicksortInt* – a typical implementation of the popular divide-and-conquer sort algorithm,
- *ImageThreshold* – a common task in many computer vision and graphics applications – algorithm to classify pixels as "dark" or "light",
- *VideoConvolute* – an algorithm that applies a convolution matrix to a portion of a video.

---

[8] Binaryen, https://github.com/WebAssembly/binaryen

We have also implemented a small benchmark test[9] that utilizes Fibonacci's *Trial Division* integer factorization algorithm in order to compute large prime numbers. The same benchmark was written in three different languages, Native C, WebAssembly, and JavaScript, to analyze how well the algorithm performs in different execution environments. In each of the cases, we decided to compute $10^7$ prime numbers.

Two versions of the *Trial Division* algorithm, namely the Native C and the JavaScript version, were implemented from scratch, while WASM binaries were produced by the compiling Native C codebase to WebAssembly utilizing Emscripten SDK. The codebase was compiled with flags that produce more optimized builds in terms of performance in order to showcase the focus of this study. The produced WASM binaries were then instantiated as modules in a JS file.

*B. Testing Environment Configuration*

The benchmark tests were obtained from our testing environment consisting of:

- clang version 11.0.3 (clang-1103.0.32.62)
- Emscripten v2.0.12
- Node v15.8.0
- Chrome Version 88.0.4324.146 x64

The operating system utilized for testing was Mac OS X 10.15.7 (Catalina) x86_64, with 2.3 GHz Quad-Core Intel Core i5 and 8 GB 2133 MHz LPDDR3 RAM.

*C. Selected Test Cases and Discussion of Results*

In order to produce respectable and qualitative results, it is not recommended nor sufficient to run benchmarks only once, since browsers tend to perform optimizations based on number of iterations. The first iterations usually take the longest. Most browsers perform optimizations on multiple levels, which happen on subsequent iterations. The browser is able to do so because of the various assumptions derived from the current execution context. However, the assumptions can be good for the end user or lead in the wrong direction. In such a case, browsers also perform so-called deoptimizations. The whole process of optimization and deoptimization[10], along with other internal factors such as garbage collection, can cause a variation in the end results. Thus, to describe the general picture, we must let the browser to "warm up".

Following this idea, our preliminary test results derived from our benchmark tests were gathered in *350* iterations. In order to produce more precise results, we discarded the first *50* runs, which would commonly occur in the warmup phase. Thus, we are left with *300* iterations, which are expressed in an average case observed in milliseconds comparing the equivalent code implementations written in WebAssembly and pure JavaScript, as shown in Fig. 2.

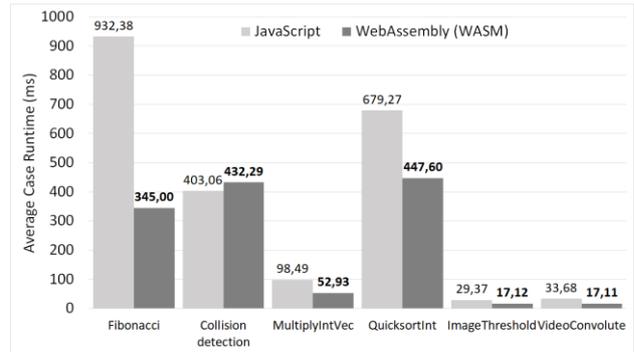

Figure 2. Preliminary benchmark test results of average case runtime (in ms) for selected benchmarks in JavaScript and WebAssembly

In almost all of the scenarios shown above, WebAssembly seems to outperform the equivalent implementations written in standard JavaScript. When running the *Fibonacci* benchmark, WebAssembly binary outperforms the equivalent JS implementation by as much as 62.97%. Similarly, the *QuickSort algorithm* written in WebAssembly performs 34.1% better in comparison to the JavaScript implementation. The *ImageThreshold* and *VideoConvolute* benchmarks show 41.7% and 49% decrease in average runtimes, respectively. However, in the case of the *Collision detection* benchmark, WebAssembly seems to be outperformed by its JavaScript counterpart. This specific benchmark seemed to perform well in the first set of iterations and has only gone downward. This could potentially be due to the quality of compiled version generated with Emscripten and the browser's optimizations not used. Nevertheless, WebAssembly demonstrated a more predictable average performance since it didn't seem to fluctuate as its JavaScript counterparts in terms of the process of optimization and deoptimization performed by the browser.

Likewise, the preliminary results illustrated in Fig. 3. showcase the custom implementation of Fibonacci's *Trial Division* integer factorization algorithm, which was performed in the same number of *350* iterations.

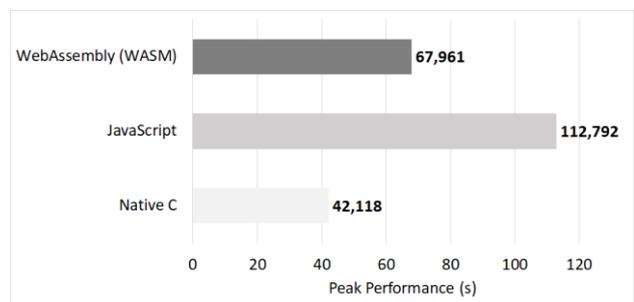

Figure 3. Peak performance in seconds obtained from a custom benchmark test featuring the implementation of Trial Divison algorithm in selected languages

Fig. 3 shows the apparent advantage of WebAssembly in memory-intensive workloads. The WASM binary outperformed the pure JavaScript implementation by as much as 39.74%, clearly demonstrating the capability of WebAssembly to achieve near-native speeds.

---

[9] JS vs WASM Fibonacci, https://takahirox.github.io/WebAssembly-benchmark/

[10] JavaScript engine fundamentals: Shapes and Inline Caches, https://mathiasbynens.be/notes/shapes-ics

## V. FUTURE WORK

Our research is a continuous project of over three years exploring different aspects of the Web-based ecosystem spanning from a stacked virtual machine, native images in the cloud, and the newest addition – the WebAssembly. Still, WebAssembly offers many more opportunities for analysis and comparison, including wasm3, a high-performance WebAssembly interpreter written in C. As our tests were performed with limited resources within an academic environment, larger scale and more extensive operations and benchmarking would suffice preciser results.

## VI. CONCLUSION

WebAssembly is practically supported by all major browser vendors, some of which are the most profitable and influential companies in the world. The current state of the system is already viable, with results achieving unexpectedly high marks. In this paper, we presented our experiences with WebAssembly and TruffleWasm within the GraalVM-based ecosystem, as well as test results with several different benchmarks and within our academic environment. We believe that further hardening and developing the system will overcome contemporary industry benchmarks. The omnipresence and the embeddability of WebAssembly in the IT ecosystem will further bolster its position as a significant player in the field.